\documentclass[aps,prc,reprint,superscriptaddress,longbibliography,nofootinbib,floatfix]{revtex4-2}

\usepackage{graphicx} % Required for inserting images
\usepackage{subcaption}
\usepackage{amsmath,amssymb}  
\usepackage{bm}  
\usepackage[pdftex,bookmarks,colorlinks,breaklinks,hidelinks]{hyperref}  
\newcommand{\nbar}{$\bar n$\;}
\newcommand{\nbarp}{$(\bar np)\;$}
\newcommand{\pbarn}{$(\bar pn)\;$}
\newcommand{\pbar}{$\bar p\;$}
\newcommand{\pbarp}{$(\bar pp)\;$}
\newcommand{\NNbar}{$\overline{\mathcal{N}}\mathcal{N}\;$}

\begin{document}
\title{The unfinished picture of low-energy antineutron interactions: open issues and hints for future research possibilities}

\author{A. Filippi}
\email{filippi@to.infn.it}
\affiliation{Istituto Nazionale di Fisica Nucleare, Sezione di Torino, Italy}

\begin{abstract}
This report examines the open questions that remain unsolved following the measurements with antineutrons ($\bar n$) as probes conducted up to the 1990s at the LEAR facility at CERN. It also presents suggestions for possible new experiments at a future, upgraded AD complex, which can potentially provide access to new areas of physics.
\end{abstract}

\maketitle
\section{Introduction}

\subsection{\texorpdfstring{$\bar n$\;} iinduced interactions and annihilation dynamics}

Using antineutrons as probes to study the dynamics of \NNbar annihilations provides several benefits over reactions initiated by antiprotons. The key point is that the \nbarp system has a definite isospin of $I=1$, while the \pbarp system contains both $I=0$ and $I=1$ components. This acts as a strong selection rule, ruling out several initial states and restricting the possible combinations of quantum numbers for any intermediate states or resonances that can be formed after the annihilation.

On the other hand, the \pbarn system, which can be realized using a deuterium target, displays the same isospin features as the \nbarp system. Antiprotons provide several practical benefits for experiments, such as the ability to undergo annihilations at rest, while \nbar annihilation must occur in flight and therefore it involves a wider spectrum of initial partial waves, which depend on the available energy. Moreover, \pbar beams can be generated with comparatively high intensities, enabling the collection of higher statistics.

However, annihilations in the \pbarn system are affected by significant complications. The neutron bound in the deuteron carries a Fermi momentum that cannot be directly determined, which prevents a complete reconstruction of the reaction kinematics. Although, in principle, the recoiling nucleon can be measured, its momentum is typically small and therefore it is hard to detect. In addition, the recoiling nucleon may rescatter either with other nucleons or with the particles generated in the annihilation, potentially degrading the cleanliness of the final state.

Antiproton beams interacting with deuteron targets face an additional challenge: annihilation can occur on either the proton or neutron, creating two distinct sources that must be carefully disentangled to isolate the pure $I=1$ component.

A key advantage of using antineutron beams as probes is the complete absence of Coulomb interactions. Unlike antiproton-proton (or nuclear) scattering, which is distorted by electromagnetic effects, $\bar{n}$ interactions are governed solely by the strong force, enabling cleaner and more precise measurements.

Further details on $\bar n$ physics are reported in Ref. \cite{re:physrep}. In the following, a short summary of the most important features of antineutron interactions will be given.

\subsubsection{Antineutron beams: historical overview}

Two primary techniques had been leveraged to produce antineutron beams: (i) extraction from external targets, where protons strike a solid target (typically Fe or Cu), and time-of-flight information is used to select the emitted $\bar{n}$, and (ii) production via the charge-exchange (CEX) reaction.

Brando \textit{et al.} first employed the external-target method at the AGS-PS in 1981, extracting antineutrons in the 0.3–1 GeV/$c$ momentum range \cite{re:brando}.

On the other hand, the $\bar p p\rightarrow \bar n n$ CEX reaction is endothermic, with a threshold momentum of 97.5 MeV/$c$, and was exploited to produced antineutrons on CH$_2$ or liquid hydrogen targets.
Gunderson \textit{et al.} pioneered this technique at Argonne ZGS in 1981 \cite{re:gunderson}, using a 1 GeV/$c$ $\bar{p}$ beam. Armstrong \textit{et al.}, later in 1987, refined this method at BNL AGS  \cite{re:armstrong}, employing 505–520 MeV/$c$ $\bar{p}$ beams on a segmented CH$_2$ target to yield a continuous $\bar{n}$ spectrum (100–500 MeV/$c$) at a  $\sim$0.2 $\bar{n}$/s rate.

A major step forward occurred at CERN's LEAR (Low Energy Antiproton Ring) in the late 1980s, exploiting intense low-energy $\bar{p}$ beams. PS178 pioneered tagged $\bar{n}$ beams in 1988 \cite{re:cugusi}, using CEX on a liquid H$_2$ target with a $10^6$ $\bar{p}$/s beam, enabling a full reaction reconstruction.

LEAR's largest $\bar{n}$-production setup was installed at PS201 (OBELIX) and in operation since 1990 \cite{re:agnello}: the apparatus consisted of a cylindrical magnetic spectrometer built around a 0.5 T dipole. An upstream production liquid H$_2$ target intercepted 305 or 412 MeV/$c$ $\bar{p}$ beams; forward-directed $\bar{n}$'s, produced via CEX,  were collimated, while other charged $\bar{p}$-annihilation products were detected and the corresponding reactions vetoed by a surrounding scintillator hodoscope (called VETO-BOX).

Due to space constraints, OBELIX could not tag $\bar{n}$'s directly. Post-collimation, the \nbar beam traversed a second liquid H$_2$ reaction target, that sometimes was followed by disk-shaped nuclear targets. Depending on the incident $\bar{p}$ momentum, the $\bar{n}$ flux intensity ranged in the interval 30–60 $\bar{n}$/10$^6$ $\bar{p}$, and the experiment could collect $\sim$35 million annihilation events over three data-taking periods.

\begin{figure}[hbt]
    \centering
    \includegraphics[width=0.95\linewidth]{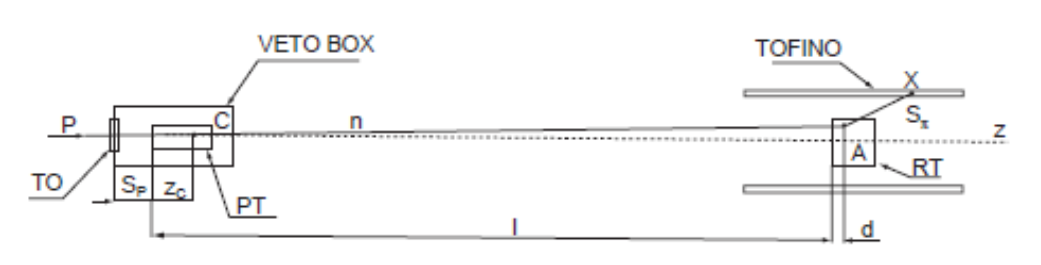}
    \caption{Sketch (not to scale) of the experimental \nbar beam arrangement of the OBELIX experiment at LEAR. From Ref. \cite{re:agnello}.}
    \label{fig:lineaNbar}
\end{figure}

The $\bar{n}$ momentum was derived iteratively from time-of-flight (TOF), as described in Ref. \cite{re:iazzi} and shortly summarized in the following. 
A sketch of the experimental arrangement is reported in Fig. \ref{fig:lineaNbar} \cite{re:physrep}. 

The total TOF, from beam scintillator $T_0$ to the first hit  of the annihilation products on a suitable scintillator hodoscope, comprises four parts (see Fig. \ref{fig:lineaNbar} for reference):
\begin{enumerate}
\item   $\bar{p}$ travel time $t_1$ for the beam scintillator ($T_0$) to the entrance window of the PT Production Target, over the $s_{\bar p}$ path length;
\item $\bar{p}$ travel time inside PT over the $z_C$  path length from the entrance target wall to the CEX vertex ($C$);
\item $\bar{n}$ time-of-flight from the CEX vertex to the $A$ annihilation vertex in the RT Reaction Target;
\item  time for the fastest pion produced in the annihilation to the $X$ hit on the scintillator hodoscope (TOFINO).
\end{enumerate}

This translates  to the following equation: 
\begin{eqnarray} 
t_{meas} & = & t_1 + t_2 + t_3 + t_4 = \nonumber \\ 
& = & \frac{s_{\bar p}}{v_{\bar p}(0)} + \int_{0}^{z_c}\frac{dz^\prime}{v_{\bar p}(z^\prime)} + \frac{l+d}{v_{\bar n}} + \frac{s_\pi}{v_\pi} 
\label{eq:time}
\end{eqnarray}

The first contribution is known from the incoming $\bar{p}$ momentum at the target entrance. The fourth can be estimated by assuming the charged pion travels at light speed ($v_\pi = c$), after reconstructing the annihilation vertex A from at least two charged tracks, to determine the distance $s_\pi$ to the hit on the hodoscope.

The critical step is to correctly determine the CEX vertex position, which affects both the second and third TOF contributions. With forward collimation making the $z$ coordinate dominant, $z_c$ yields the $\bar{p}$ travel time $t_2$ inside the target and the $\bar{n}$ flight time $t_3$ over the distance $l$ to the annihilation vertex. The $\bar{n}$ speed $v_{\bar{n}}$ follows from two-body CEX kinematics once the \nbar emission angle is known. 

$z_c$ emerges iteratively, discretizing the  $p_{\bar{p}}(z)$ momentum along the target to model the energy loss. The equation

\begin{equation} z_c = \int_{ p_{\bar p(0)}}^{p_{\bar p}}\frac{\beta\, dp}{dE/dx}
\label{eq:zeta}
\end{equation}
is solved iteratively for the $z_c$ value that matches the measured total TOF to the sum of the mentioned four components (Eq.~\ref{eq:time}); in Eq. (\ref{eq:zeta}) $\beta$ is $\bar{p}$ velocity and $dE/dx$ is energy loss per path length in the target.

These iterations introduce a relative uncertainty in $\bar{n}$ momentum, which scales with the step size and rises at low momenta. Thus, a momentum determination for $\bar{n}$'s below 50 MeV/$c$ was deemed unreliable and the corresponding events were cut off.

\begin{figure}
    \centering
    \includegraphics[width=0.95\linewidth]{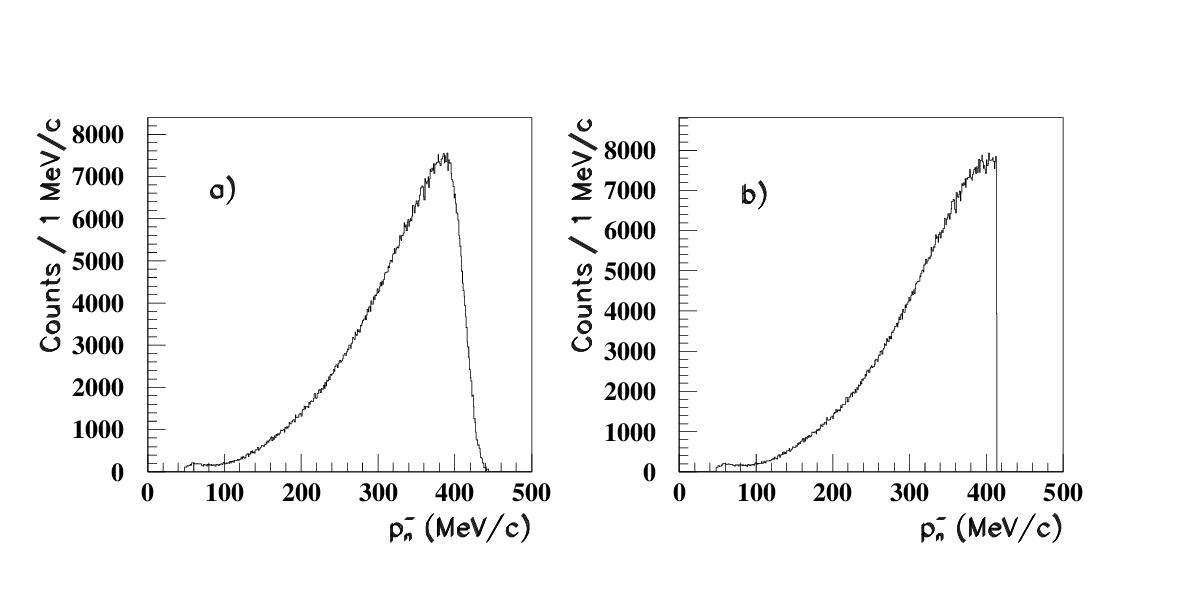}
    \caption{Experimental momentum spectrum of antineutrons produced via CEX reaction in the OBELIX \nbar facility. From Ref. \cite{re:physrep}. The plot on the left is a raw spectrum, including random coincidences without  background subtraction, while on the right the true spectrum is depicted after the
    corrections for apparatus resolution.}
    \label{fig:nbarMomentum}
\end{figure}

Typical antineutron momentum spectra as reconstructed in OBELIX are reported in Fig. \ref{fig:nbarMomentum}. 

\subsubsection{
\texorpdfstring{$\bar n p$\;} ttotal cross section} 
\label{total}

\begin{figure}
    \centering
    \includegraphics[width=0.8\linewidth]{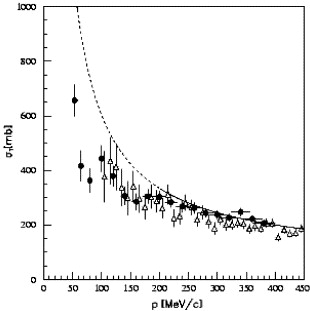}
    \caption{Comparison of \NNbar total cross section at low momenta. 
    The open triangles are from Ref. \cite{re:iazzi}, the full points are from Ref. \cite{re:armstrong}. The solid line, whose extrapolation to lower momenta is shown
    by the dashed part, is from Ref. \cite{re:bugg}. Figure from Ref. \cite{re:physrep}.}
    \label{fig:totalxsec}
\end{figure}

The total cross sections for antinucleon-induced interactions below 500 MeV/$c$ are among the best-measured quantities in antinucleon physics. Fig.~\ref{fig:totalxsec} collects data from Armstrong \textit{et al.} \cite{re:armstrong} (open triangles) and OBELIX \cite{re:iazzi} (black points).

Armstrong \textit{et al.} used a transmission technique (empty vs. full target), revealing a trend compatible with $A + B/p$.
OBELIX also measured total cross sections in the 50–400 MeV/$c$ range via transmission with a thick target and narrow beam (total error $\sim$10\%); results overlap well with those by
Armstrong \textit{et al.}

Fig.~\ref{fig:totalxsec} includes, for completeness, Bugg \textit{et al.}'s solid/dashed extrapolation for $\bar{p}p$ \cite{re:bugg}. Notably, $\bar{n}p$ cross section lies below $\sigma_T(\bar{p}p)$ under 200 MeV/$c$.

\subsubsection{
\texorpdfstring{$\bar n p$\;} aannihilation cross section.}
\label{annih}

The first $\bar{n}p$ annihilation cross section measurements in the 500–800 MeV/$c$ range came from Banerjee \textit{et al.}'s bubble-chamber experiment (1985) \cite{re:banerjee1,re:banerjee2}, using CEX-produced $\bar{n}$; they reported as integrated cross section $\sigma_{ann} = (55.4 \pm 2.2)$ mb.

Armstrong \textit{et al.} derived $\sigma_{ann}$ by subtracting a parametrization of the elastic cross section from the total one (procedure affected by a 15–20\% uncertainty); the result was consistent with a $A + B/p$ scaling law. 

At BNL AGS, Mutchler \textit{et al.} (1988) \cite{re:mutchler} measured $\beta\sigma_{ann} = (40 \pm 3)$ mb at 22 MeV/$c$ and $(32 \pm 5)$ mb at 43 MeV/$c$, as well as, for the first time, the $S$-wave scattering length imaginary part $a_I = (-0.83 \pm 0.07)$ fm.

\begin{figure}
    \centering
    \includegraphics[width=0.8\linewidth]{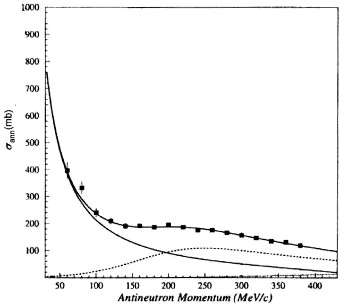}
        \caption{Fit of $\sigma_{ann}(\bar np)$ measured by OBELIX up to 400 MeV/$c$ including $S$- (solid line), $P$- (dotted line) and $D$- (dashed line at the bottom of the picture) wave contribution. From Ref. \cite{re:bertin}.}
    \label{fig:annOBX}
\end{figure}

OBELIX, later on, covered the 50–400 MeV/$c$ range (see Fig.~\ref{fig:annOBX}) \cite{re:bertin}.
The points measured by OBELIX agree well with Armstrong data but carry a normalization error as large as 7\%. They can be fitted with an effective-range expansion (including $S$-, $P$-, $D$-waves), with $D$-wave contributing $(4.7 \pm 0.6)\%$ up to 200 MeV/$c$.

\subsubsection{
\texorpdfstring{$\bar n$}--nucleus annihilation cross section.}
\label{nuclei}

OBELIX sometimes used additional nuclear targets (made of various materials) downstream of the liquid H$_2$ production target, to measure $\bar{n}$ annihilation cross sections on different nuclei as a function of projectile momentum \cite{re:astrua}.

\begin{figure}
    \centering
    \includegraphics[width=0.8\linewidth]{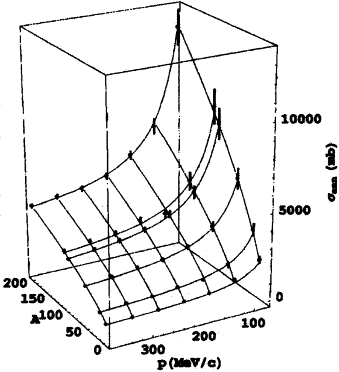}
    \caption{Three-dimensional plot of the measured $\sigma_{ann}(p_{\bar n};\; A)$, as a function of the antineutron momentum  and the atomic number. From Ref. \cite{re:astrua}.}
    \label{fig:sigmaAnnNuclei}
\end{figure}

An $A^{2/3}$ scaling with the $A$ nucleus mass number emerged (2\% accuracy), implying surface-dominated annihilations with localized, large hadronic cross sections. The momentum dependence roughly followed the antineutron $\beta$; the annihilation cross sections were overall consistent with $\sigma_{ann} = (\alpha + B/p) A^x$ (see Fig.~\ref{fig:sigmaAnnNuclei}) trend, with $x\sim 2/3$.

\section{Open puzzles in Antineutron Behavior}

LEAR data analyses performed up to early 2000s reveal puzzles when compared with later AD experiments and modern theory. Key open questions are summarized below.

\subsection{\texorpdfstring{\nbar\;}  aannihilation cross sections on nuclei at low momenta}

\begin{figure}
    \centering
    \includegraphics[width=\linewidth]{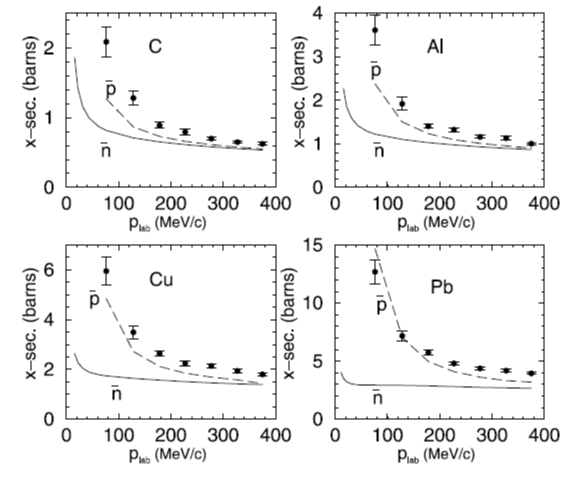}
    \caption{Total annihilation cross sections compared to
    the results from the model by Friedman \textit{et al.}
    (solid curves for $\bar n$, dashed curves for $\bar p$.
    From Ref. \cite{re:friedman}. Experimental data for
    $\bar n$ from Ref. \cite{re:astrua}.}
    \label{fig:friedmanModel}
\end{figure}

Friedman \textit{et al.} (2014) \cite{re:friedman} developed an optical-potential model fitting the
momentum dependence of $\bar{p}A$ annihilation cross sections on nuclei. However, the potential fails to reproduce $\bar{n}$-induced interactions: as Fig.~\ref{fig:friedmanModel} shows, predictions (solid line) systematically underestimate the data across different nuclear species, pointing at a missing component in the model specific for $\bar{n}$ interaction.

\begin{figure}
    \centering
    \includegraphics[width=0.9\linewidth]{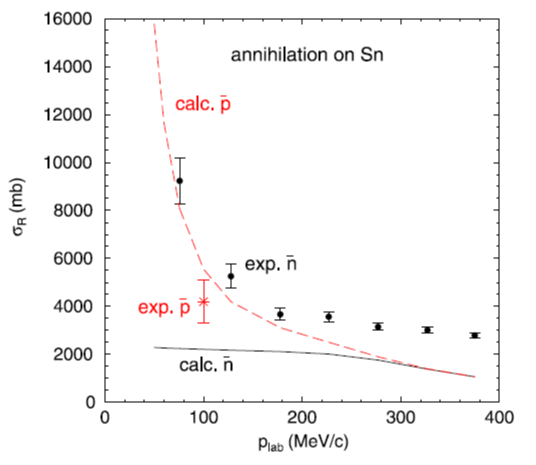}
    \caption{Total $\overline{\mathcal N}$ annihilation cross sections on Sn 
    compared to the results based on the potential derived in Ref.\cite{re:friedman} (solid black curve for $\bar n$, dashed red curve for $\bar p$). Experimental points for $\bar n$
    from Ref. \cite{re:astrua}, for $\bar p$ (single red point) from \cite{re:asacusaSN}. From Ref. \cite{re:friedman}.}
    \label{fig:annOnSn}
\end{figure}

Indeed, antineutrons lack the Coulomb component essential for $\bar{p}$, and this suggests that a focusing effect could be missing. The scarce $\bar{p}A$ existing data prevent comparison: ASACUSA's unique  point for \pbar annihilation on Tin \cite{re:asacusaSN} is exceeded both by the model and by the $\bar{n}$-Sn cross section measurements (see Fig.~\ref{fig:annOnSn}).

Such discrepancies demand precise, simultaneous measurements of in-flight $\bar{n}$ and $\bar{p}$ annihilation cross sections, ideally on identical targets and energies, using the same experimental setup.

\subsection{The elastic \texorpdfstring{$\bar n p$\;} ccross section}

\begin{figure}[hbt]
    \centering
    \includegraphics[width=0.8\linewidth]{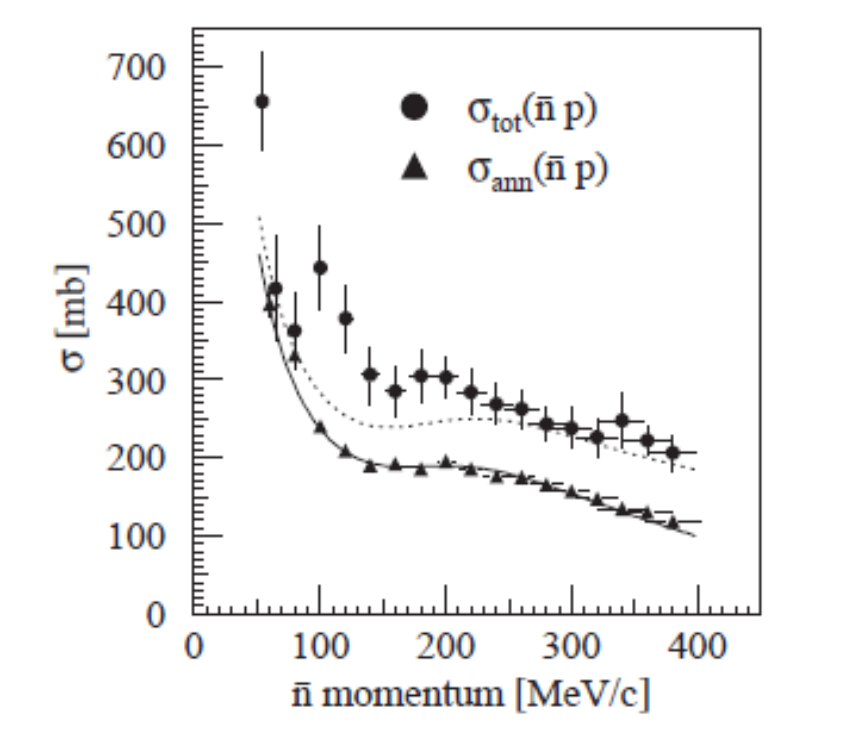}
    \caption{Experimental value of the total (Ref. \cite{re:iazzi}, black triangles) and annihilation (Ref. \cite{re:bertin}, black circles) \nbarp cross sections. Data by OBELIX. The solid curve represents the calculation of $\sigma_{ann}(\bar np)$  performed in Ref. \cite{re:bertin}, the dotted one the calculation of $\sigma_T(\bar np)$ by using the same parametrization.}
    \label{fig:dipbump}
\end{figure}

OBELIX derived the $\bar{n}p$ elastic cross section by subtracting the annihilation cross section from the total one. As detailed in Sec.~\ref{annih}, the annihilation trend with the momentum is smooth, yet the total cross section (Sec.~\ref{total}) shows a dip-bump trend at 65–80 MeV/$c$. as shown in Fig.~\ref{fig:dipbump}.

This dip-bump behavior prevents fitting both cross sections with shared effective-range parameters \cite{re:mahalanabis}; $\sigma_T$ fits fail markedly.

\begin{figure}
    \centering
    \includegraphics[width=0.8\linewidth]{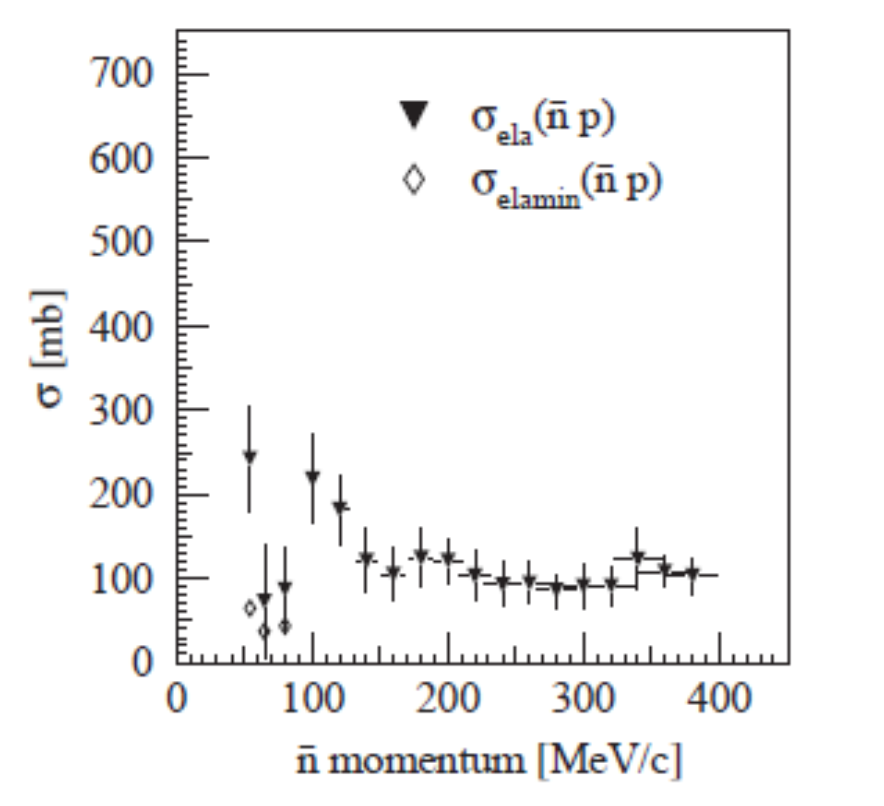}
    \caption{Black triangles: elastic cross section as a function of $p_{\bar n}$ obtained from the difference between the total \cite{re:iazzi} and annihilation \cite{re:bertin} cross sections. Diamonds: lower limits of the elastic cross sections due to the unitarity constraints. From Ref.\cite{re:iazzi}.}
    \label{fig:elastic}
\end{figure}

The origin of such an anomaly remains unclear, possibly due to an irregular elastic behavior (see Fig. \ref{fig:elastic}, post-subtraction). Points at 64.5 and 80 MeV/$c$ the $S$-wave meet the unitarity lower bound: 
$\sigma_{el} \geq k^2/(4\pi) \cdot \sigma_T^2$.

Proposed explanations invoke a near-threshold quasi-nuclear bound state in the $\bar{n}p$ spin-triplet configuration \cite{re:kudryatsev}. Strong isospin violation between $I=0$ and $I=1$ channels could determine different low-momentum \NNbar behaviors: large, opposite-sign scattering lengths $a_0$ and $a_1$ might manifest in a repulsive interaction in $\bar{p}p$, while an attractive one should be effective in $\bar{n}p$.

Alternatively, elementary nuclear mechanisms, like a Ramsauer-Townsend effect \cite{re:ramsauer}, could suppress the elastic cross section via an interference mechanism.

\subsection{\texorpdfstring{$I=0$\;} vvs. \texorpdfstring{$I=1$\;} aannihilation sources}

The isospin ratio $R$ may quantitatively be defined as
$$ R = \frac{\sigma_T(I=0) + \sigma_T(I=1)}{2\sigma_T(I=1)}$$
or, alternatively, it can be expressed by  $$r = \sigma_T(I=0)/\sigma_T(I=1) = 2R - 1$$. 

At 70 MeV/$c$, $r=2.5 \pm 0.4$ indicates $I=0$ dominance (to be compared to $r=1.1 \pm 0.1$ at 300 MeV/$c$), possibly generated by coherent $\omega$, $\delta$, $\rho$-exchange (central/tensor terms in the medium range \NNbar interaction) \cite{re:dover}.

\begin{figure}
    \centering
\includegraphics[width=0.95\linewidth]{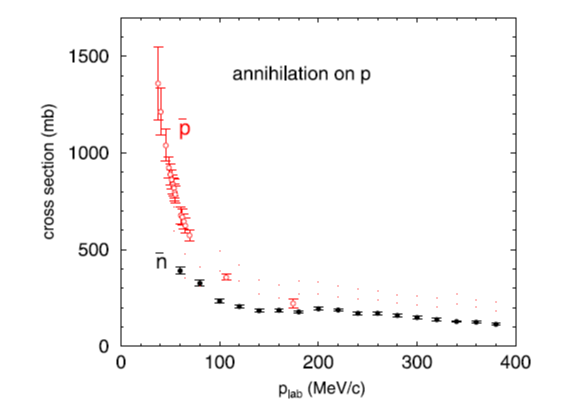}
    \caption{Total \nbarp annihilation cross section (full black points) and \pbarp annihilation cross sections (red markers), from Ref. \cite{re:friedman}.}
    \label{fig:comparisonPbarNbar}
\end{figure}

At low momenta, $\sigma_{ann}(\bar{n}p) < \sigma_{ann}(\bar{p}p)$ consistently \cite{re:zenoni} (as shown in Fig.~\ref{fig:comparisonPbarNbar}): the value $R \sim 1.7$ ($r=2.4 \pm 0.4$) is obtained at 70 MeV/$c$, while it approaches 1 at 300–500 MeV/$c$, and then it rises again (with $r \to 1.5$) near 700 MeV/$c$. In addition, the Bombay-CERN-Neuchâtel-Tokyo data \cite{re:banerjee1}, dating back to the 70's, report $\sigma_{ann}(\bar{n}p)=55.4 \pm 2.2$ mb, to be compared to $\sigma_{ann}(\bar{p}p)=77.9 \pm 0.6$ mb (500–800 MeV/$c$), and signal a further inconsistency which calls for deeper scrutiny.

\subsection{The differential CEX cross sections}

\begin{figure}[hbt]
    \centering
    \includegraphics[width=0.8\linewidth]{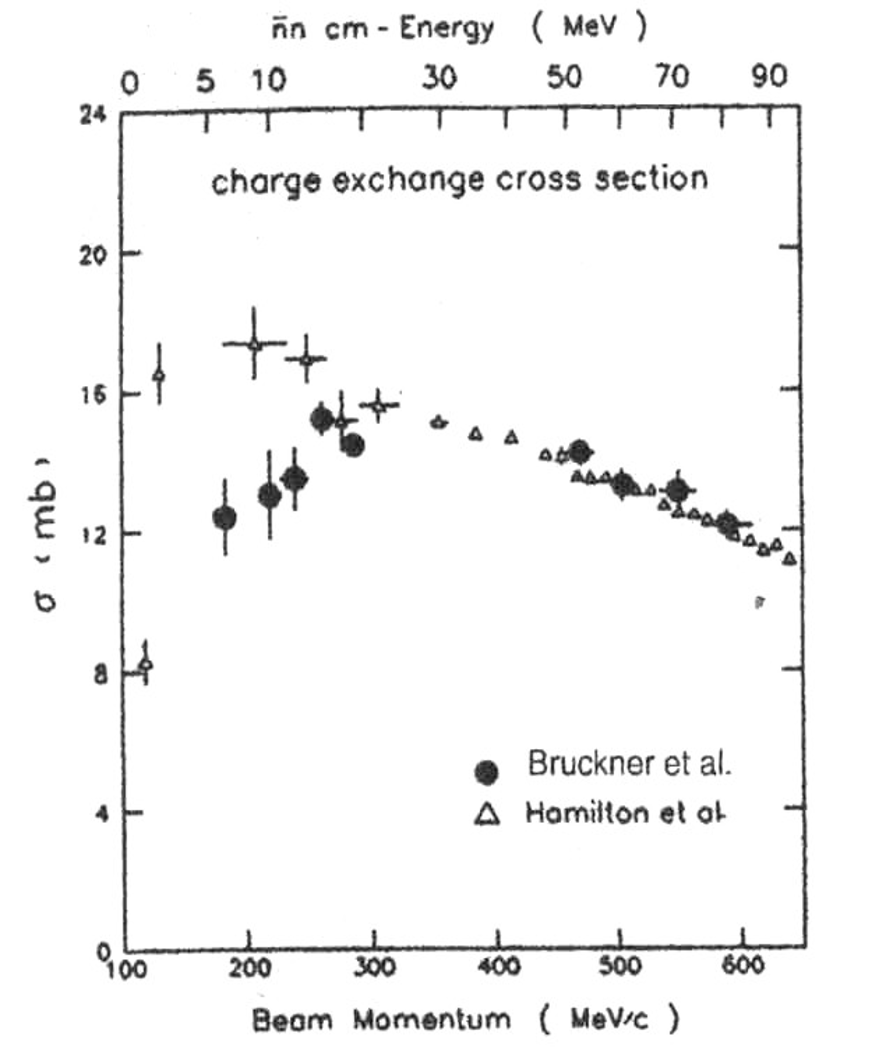}
    \caption{CEX cross sections data: black circles by Hamilton \textit{et al.} \cite{re:hamilton}, open circles by Br\"uckner \textit{et al.} \cite{re:bruckner}.}
    \label{fig:cex}
\end{figure}

Charge-exchange (CEX) processes also exhibit unresolved issues: few low-momentum measurements exist, which present strong disagreements (Fig.~\ref{fig:cex}, all data from experiment performed in the Eighties).

Near-threshold data, in fact, disagree sharply: Hamilton \textit{et al.}'s data (1980, open triangles) \cite{re:hamilton}  follow the endothermic trend, while Br\"uckner \textit{et al.}'s data (1987, full points) \cite{re:bruckner}  show a linear decrease, which is still unexplained and warrants new measurements.

\begin{figure}[bht]
    \centering
    \includegraphics[width=0.9\linewidth]{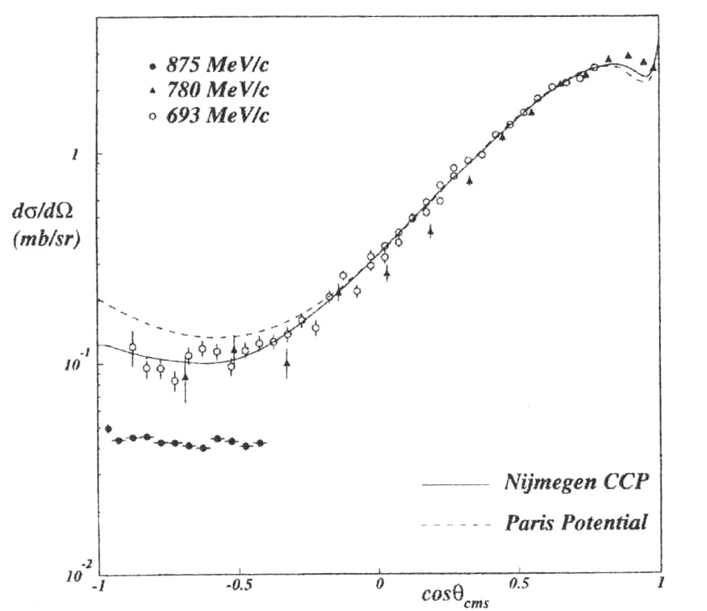}
    \caption{Differential Charge-EXchange cross section. Open and full circles are
    from Ref. \cite{re:ps199}, full triangles are from Ref. \cite{re:nakamura}.}
    \label{fig:differentialCEX}
\end{figure}

Differential CEX is even murkier, with sparse high-momentum data. Figure \ref{fig:differentialCEX} compiles full-angular scans: PS199 (1985) \cite{re:ps199} measurements at 693 MeV/$c$ (open circles) and 875 MeV/$c$ (full circles) are shown together with Nakamura \textit{et al.}'s data (1987) \cite{re:nakamura} at 780 MeV/$c$ (full triangles).

PS199 data diverge sharply from others, especially at backward angles and zero degrees.

\begin{figure}
    \centering
    \includegraphics[width=0.8\linewidth]{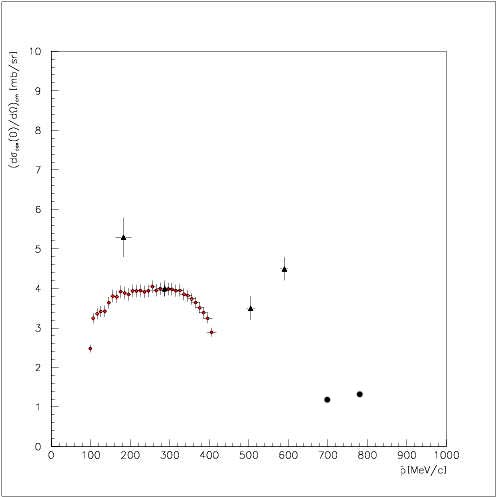}
    \caption{Forward Charge-EXchange cross section. The triangles are by Br\"uckner \cite{re:bruckner} , the red points from 100 to 400 MeV/$c$ are by OBELIX \cite{re:iazzi}, the full points at $\sim 800$ MeV/$c$ are from Nakamura \cite{re:nakamura}.}
    \label{fig:0degCEX}
\end{figure}

Forward discrepancies persist at lower momenta: Br\"uckner \textit{et al.}'s data (1987)   at 183, 287, 505, 590 MeV/$c$  \cite{re:bruckner}  vs. indirect backward estimates by OBELIX  (99–400 MeV/$c$) \cite{re:iazzi} are shown in Fig.~\ref{fig:0degCEX}.
OBELIX aligns with Br\"uckner's data near 300 MeV/$c$ and follows the endothermic trend of Hamilton's measurements  (Fig.~\ref{fig:cex}), but the two experiment diverge sharply at 183 MeV/$c$. New precise differential CEX measurements especially at forward angles are therefore essential.

\section{Additional dynamic-related puzzles}

Several effects observed by OBELIX in $\bar{n}$ annihilations to exclusive final states, characterized by flavor signatures, are tied to annihilation dynamics mechanisms  and final state quark content, and are mainly to be investigated through spectroscopic techniques. Some of these phenomena still lack full explanations, which demand more precise experimental data and refined models for a complete accounting.

\subsection{Dynamical selection rules and the onset of strangeness in annihilation reactions}

OBELIX probed $\bar{n}$-induced annihilation to two-body states with distinct flavor content, to trace the strangeness production onset in annihilation processes.

Low-momentum, or at rest, annihilations can ideally access non-perturbative nucleon sea quarks.
Interestingly, non-trivial dynamical selection rules emerge, transcending quantum-number conservation and exhibiting peculiar behaviors which vary with reaction energy, \textit{i.e.} they depend on the excited $\bar{n}p$ partial waves.

\begin{figure}
    \centering
    \includegraphics[width=1\linewidth]{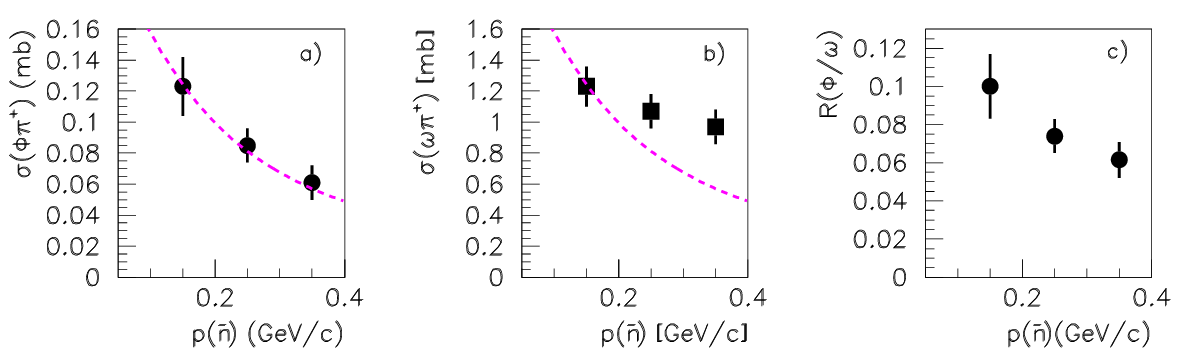}
    \caption{Trends as a function of \nbar momentum of: (a) $\bar n p\rightarrow \phi\pi^+$ and (b) $\bar n p\rightarrow \omega\pi^+$ cross sections. The curve superimposed is the trend expected for pure $S$-wave annihilation as given by Dover-Richard model \cite{re:dover}. (c) Ratio between the $\phi\pi^+$ and and $\omega\pi^+$ yields, as a function of $p_{\bar n}$. From Ref. \cite{re:phiOmega}.}
    \label{fig:ozi_phiOmega}
\end{figure}

From OBELIX data a strong contrast emerges in the production of hidden strangeness versus non-strange mesons: in $\bar{n}p \to \phi \pi^+$ and $\bar{n}p \to \omega \pi^+$ reactions, the trend of cross sections vs. momentum reveals a strong OZI violation depending on the reaction energy \cite{re:phiOmega}. $\phi \pi^+$ production follows $S$-wave annihilation (dashed line in Fig.~\ref{fig:ozi_phiOmega}), differently from the strangeness-free $\omega \pi^+$; phase-space-corrected ratios scale as well with the momentum. A strong OZI violation, up to a factor of 20 larger, indicates the nucleon strangeness content as a possible origin of
the abundant production of $\phi$ mesons: vector-meson nonet ideal mixing would preclude the presence of strange valence quarks
were they not already available, in some form, in the initial state.

This implies that $\phi$ mesons likely arise from polarized excitations of the  $\bar{s}s$ sea in nucleons and antinucleons, modulated by annihilation partial waves and quantum numbers. This mechanism echoes high-energy deep inelastic scattering's color transparency effect, where $\bar{s}s$ spins were found to align with nucleon spin.

\begin{figure}
    \centering
    \includegraphics[trim=9.35cm 0 0 0, clip, width=0.8\linewidth]
    {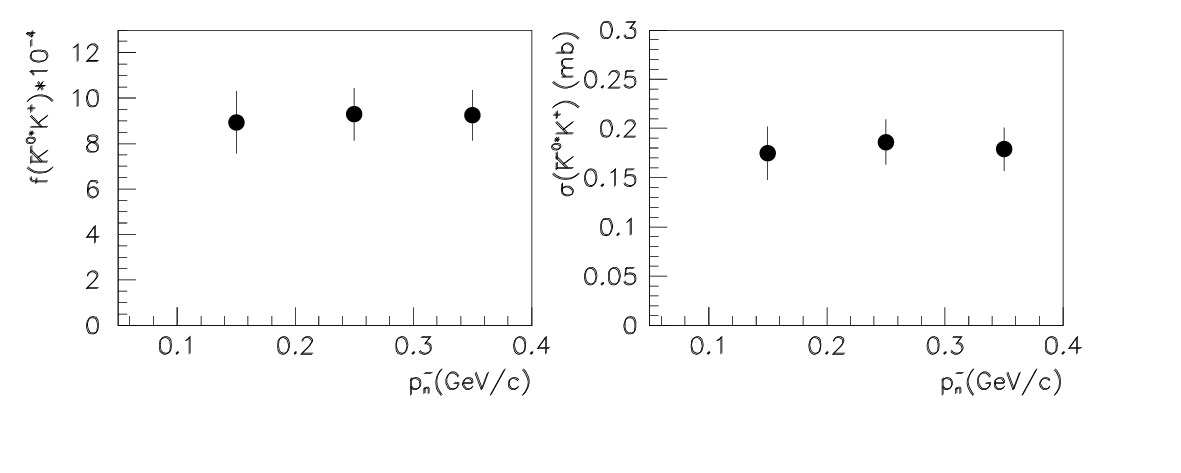}
    \caption{Trends of the cross section as a function of \nbar momentum of $\bar n p\rightarrow \overline{K^{0\ast}}K^+$. Data from Ref. \cite{re:phiOmega}.}
    \label{fig:kstar}
\end{figure}

Open-strangeness channels like $\bar{n}p \to K^{*0}K^+$, on the other hand, show no $S$-wave scaling (Fig.~\ref{fig:kstar}) \cite{re:phiOmega}, favoring the simple rescattering mechanism over more complex dynamics.

Beyond polarized sea quarks or rescattering, intermediate $\bar{s}\bar{q}sq$ strange tetraquarks can moreover play as an alternate viable hypotheses.

\begin{figure}
    \centering
    \includegraphics[trim=9.35cm 0 0 0, clip, width=0.8\linewidth]{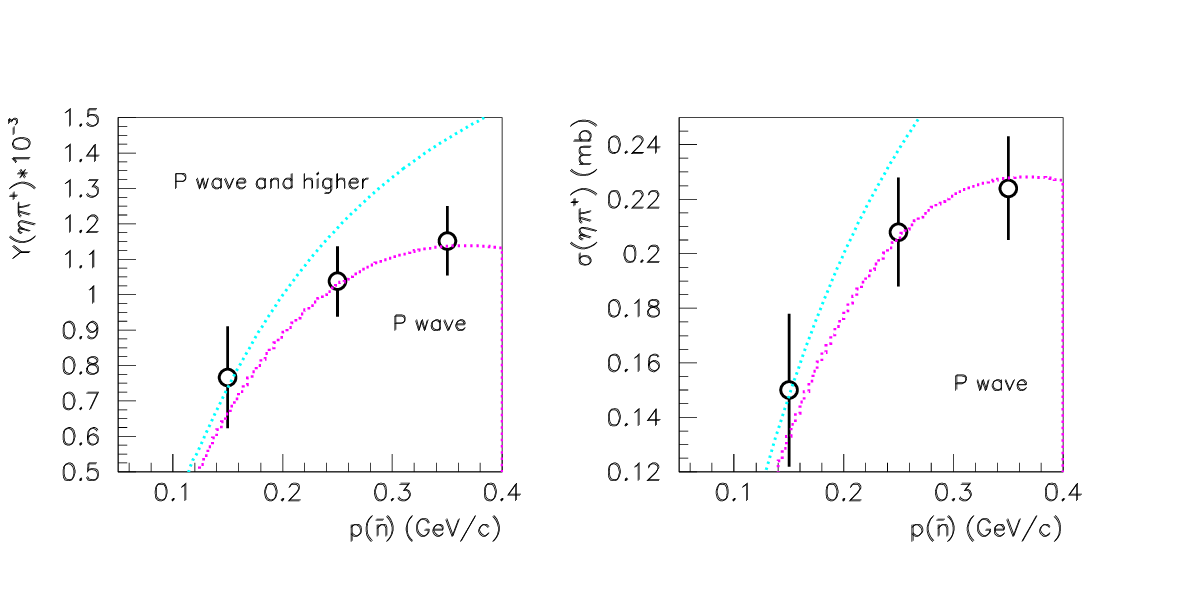}
    \caption{Trend as a function of \nbar momentum of $\bar n p\rightarrow \eta\pi^+$ cross sections. The curve superimposed are the trend expected for pure $P$-wave annihilation (magenta) and $P$ plus higher waves as provided by Dover-Richard model \cite{re:dover}. Data from Ref. \cite{re:eta}.}
    \label{fig:eta}
\end{figure}

However, other two-body channels like $\bar{n}p \to \eta\pi^+$ obey simple quark-line rules without evident anomalies \cite{re:eta}. Its cross section tracks $P$-wave momentum dependence (see Fig.~\ref{fig:eta}), matching the expected quantum-number conservation rules for the production of two pseudoscalar mesons.

\subsection{Hunting for Baryonium Signatures}

Several theoretical models predict \NNbar bound states (baryonium), based on the short-range attraction in $I=0$ ($\omega$, $\sigma$ exchange) overtaking $I=1$ repulsion (determine by $\rho$ exchange) \cite{re:dover}. Yet, the annihilation process does not provide unambiguous signatures so far, demanding more precise data and refined potentials.

In spite of substantially negative findings gathered over the years, hints for such a structure still persist, and the search is still ongoing. In the past decades, a $J^{PC}=1^{--}$ resonance (1.911 GeV/$c^2$, 29 MeV width) was suggested by E687 in six-pion photoproduction \cite{re:e687}, and a signal was observed as well by DM2 near-threshold, in $e^+e^-$ annihilation to six pions \cite{re:dm2}.
OBELIX sought for similar signatures in $\bar{n}p \to 3\pi^+ 2\pi^- \pi^0$ but found no signal, setting for such a state $\sigma \leq 0.5$ mb \cite{re:formation}.

More recently, a $X(1835)$ state close to the \NNbar  threshold was observed in high energy experiments in $J/\psi$ radiative decays, as well as 
in $B$ decays. The $X(1835)$ is a light pseudoscalar resonance first observed by BES in 2005 and confirmed by BESIII with high significance: BES observed it in 
$J/\psi \to \gamma \bar{p}p$ \cite{re:BES}, while more recently BESIII confirmed its presence in $J/\psi \to \gamma \pi^+ \pi^-\eta^\prime$ \cite{re:BES3} and $J/\psi \to \gamma 3\pi^+ 3\pi^-$ \cite{re:BES3_6pi}. Its mass $(1836.5\pm 3.0(stat) ^{+5.6}_{-2.1}(syst)$ MeV/$c^2$) and  width $(190 \pm 9(stat) ^{+3.8}_{-3.6} (syst)$ MeV) were measured, with $>20\sigma$ significance;
observed decay modes  include the also $\eta K^0_S K^0_S$ and $\gamma\phi$. High precision BESIII data are still being investigated to provide additional indications through line shapes and couplings analyses, especially close the $p\bar{p}$ threshold.
Alternative explanations for such a signal suggest the identification as
glueball, $\eta^\prime$  or $s\bar{s}$ radial excitations, but no consensus exists yet on its nature.

In addition, Belle observed a structure in $B\rightarrow D\bar{p}p$ and $B \rightarrow K\bar{p}p$ decays \cite{re:belle1,re:belle2}. While LHCb and Belle II did not provide any significant new observation for such a structure, 
investigations are still on-going to ascertain possible near-threshold dynamics related effects.

\subsection{Meson spectroscopy in gluon rich environments}

\NNbar annihilation reactions create gluon-rich environments, ideal for producing and observing exotic states, like purely gluonic glueballs or quark-gluon hybrids. 

Meson spectroscopy studies at LEAR exploited mainly antiproton annihilations at rest on hydrogen targets of different densities to select the partial wave composition of the initial $\bar{p}p$ state. Such a control is made possible by the Stark effect, a density-dependent mechanism which allows to tune the quantum numbers of the atomic levels from which annihilation occurs.

However, compared to modern experiments, LEAR's statistics remain limited, so additional \NNbar annihilation data would complement investigations of specific meson and baryon excitations.

Antineutron beams, in particular, remained underutilized for meson spectroscopy. Beyond the pioneering OBELIX experiment, which analyzed some selected and exceptionally clean exclusive channels with charged pions in the final state, with a background level below 5\%, their full potential is still to be exploited. In the analysis of the $\bar{n}p \rightarrow 2\pi^+ \pi^-$ channel, OBELIX extracted signals possibly linked to the $f_0(1500)$ and $f_2(1565)$ resonances, using 35,118 events with $\bar{n}$ momenta over the 50--300 MeV/$c$ range ~\cite{re:obx_3pi}.

\subsubsection{Channels with open and hidden strangeness}

The spectroscopy of states produced in $\bar{n}p$ annihilation with open or hidden strangeness would yield valuable insights, thanks to distinctive sets of selection rules. Although OBELIX statistics for exclusive final states with strangeness was so small as to preclude full spin-parity analyses, invariant mass distributions still reveal interesting key indications, as detailed below. Certain channels offer fresh perspectives on unconfirmed states, aided by exceptionally clean data samples.

\paragraph{$\bar{n}p \rightarrow K^+ K^- \pi^+$.}

This channel probes hidden-strangeness strangeonia: many expected states in the
$1^{--}$ sector are still unconfirmed, while some observation play as potential exotics decaying to strange quarks. Resonances decaying to $K^+K^-$ carry $J^{PC} = (even)^{++}$ or $(odd)^{--}$ quantum numbers, and include $f_0$, $f_2$, $a_0$, $\phi$ mesons and their radial excitations. Such a reaction can also illuminate open-strangeness radial excitations production, like $K^\ast$, $K_1$, and $K_2$ states.

\begin{figure}
    \centering
    \includegraphics[width=\linewidth]{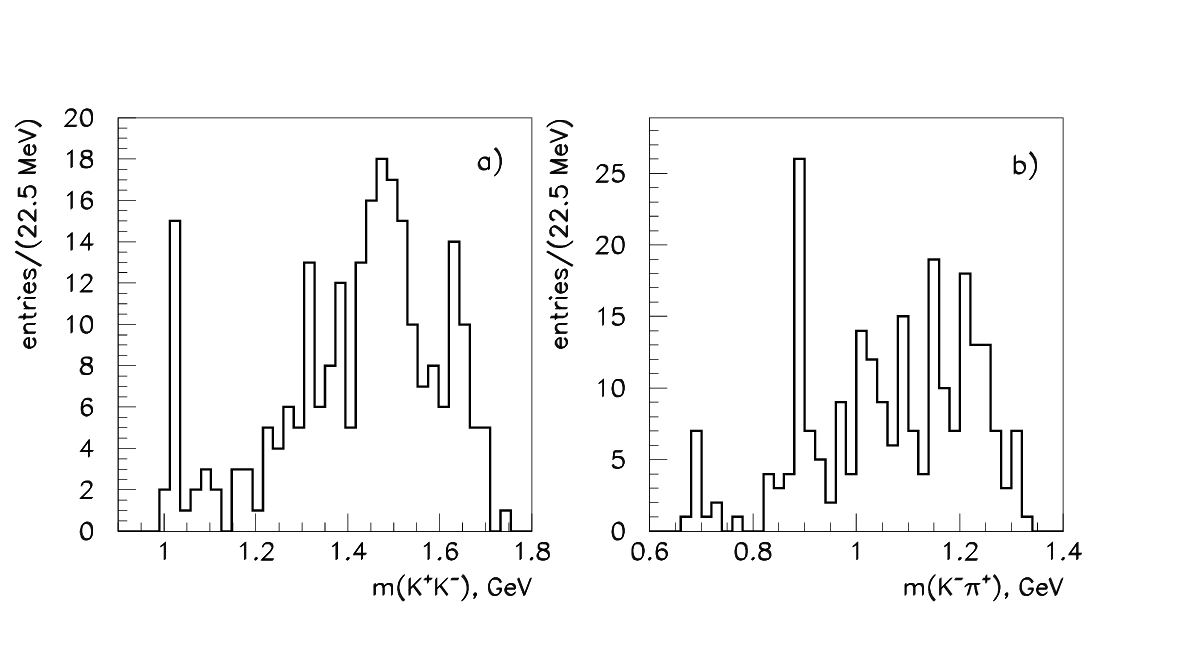}
    \caption{(Invariant mass spectra of the (a) $(K^+ K^-)$ and (b) $(K^-\pi^+)$ pairs from the $\bar np\rightarrow K^+K^-\pi^+$ reaction, with particles selected by means of a 4C fit and particle identification criteria for both kaons; OBELIX data.}
\label{fig:kkbarpi}
\end{figure}

OBELIX selected 241 exclusive events in this reaction via kaon identification (for both of them) with $dE/dx$ and $\beta$ measurements, kinematic cuts, and a 4C fit.
Figure~\ref{fig:kkbarpi}(a) displays the invariant mass distributions for the $(K^+K^-)$ system, where clear $\phi(1020)$ and $f_2(1525)$ signals appear. 
The contributions of intermediate states to the remaining spectral regions require thorough investigation, which is prevented with the available tiny statistics.

On the right panel of Fig. \ref{fig:kkbarpi}  the $(K^-\pi^+)$ invariant mass spectrum is shown, which reveals a prominent $K^{*0}(892)$ peak.

\paragraph{$\bar np\rightarrow K^0_SK^0_S\pi^+.$} 

The $(K^0_S K^0_S)$ system has fixed quantum numbers $J^{PC} = (even)^{++}$. Consequently, $\phi$ mesons, as well as, in general, other strangeonium excitations (with $J^{PC} = 1^{--}$), cannot decay into this channel. Intermediate $f_0$ and $f_2$ mesons decaying to $K^0_SK^0_S$ can pair with a pion in the final state, but this occurs only from initial states with $G = -1$ (\textit{i.e.}, at low energies: $^1S_0$, $^3P_1$, $^3P_2$ waves). Conversely, the $a_0\pi$ final state requires $G = +1$.

\begin{figure}[hbt]
    \centering
    \includegraphics[width=\linewidth]{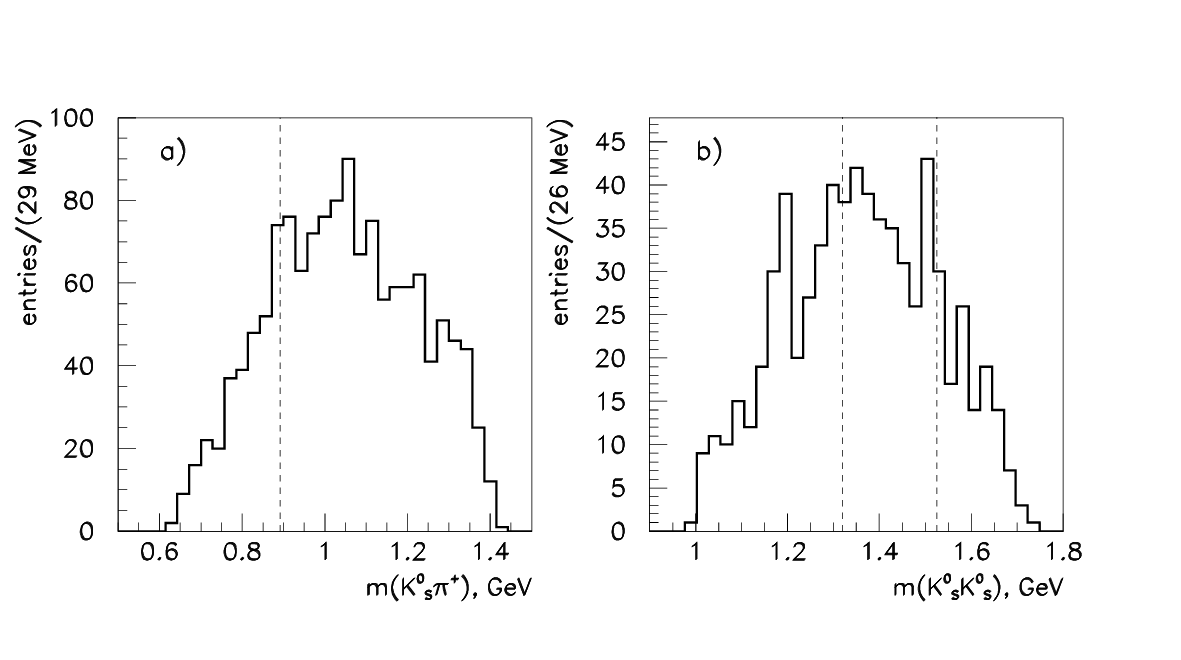}
    \caption{Invariant mass distributions of the (a) ($K^0_S \pi^+$) and (b) 
    of the
    ($K^0_S K^0_S$) system from the $\bar np\rightarrow K^0_S K^0_S\pi^+$ reaction  selected by means of a 6C fit; OBELIX data. The dashed lines in panel (b)  indicate the peak position of the
    $a_2(1320)$ and the $f_2(1525)$ resonances.}
    \label{fig:kshortkshort}
\end{figure}

OBELIX analyzed this channel, selecting 687 events for $\bar{n}$ momenta of 50--400 MeV/$c$. Although enhanced particle identification and a precise $K^0_S$ vertex reconstruction were unavailable, the $(K^0_S K^0_S)$ invariant mass distribution in Fig.~\ref{fig:kshortkshort}(b) shows preliminary indications for $a_2(1320)$ and $f_2(1525)$ signals. In the same Figure, panel (a) reports the distribution of the $(K^0_S\pi^+)$ system, where a hint for 
a $K^{\ast +}$ can be observed.

\paragraph{$\bar np\rightarrow K^0_L K^\pm\pi^\mp \pi^+.$}

This channel offers a promising avenue for inspecting resonant ($\overline{K}K\pi$) states, that would potentially add information for the resolution of  the long-standing $E/\iota$ puzzle~\cite{re:eiota}, clarifying pseudoscalar-axial-vector superpositions in the 1.3--1.5 GeV/$c^2$ region. In $\bar{n}p$ annihilation, $G$-parity conservation restricts the possible initial states for this reaction to $P$-wave, favoring axial-vector meson production.
A selection rule further constrains the $(K^0K^\pm)$ system to $I^G = 1^+$, allowing intermediate contributions to the recoiling $(\pi\pi)$ channel only from $a_0$, $a_2$, and $\rho$ resonances.

\subsection{\texorpdfstring{$\bar n-n$\;} ooscillation studies}

Searches for $\bar{n}$-$n$ oscillations probe physics beyond the Standard Model, particularly Grand Unified Theory (GUT) predictions of varying timescales. Such oscillations would violate baryon number by $\Delta B = 2$, providing unambiguous new physics signals.

Experiments dedicated to $\bar{n}$-$n$ oscillations studies divide into free-neutron and nuclear-bound searches. Concerning the latter, Super-Kamiokande set a lower limit on the oscillation period of $4.7 \times 10^8$ s (90\% CL) studying interactions on $^{16}$O \cite{re:oscillationsSKam}. Conversely, for free neutrons, which offer cleaner, model-independent measurements, the ILL experiment (1994) obtained $\tau_{\bar{n}n} > 8.6 \times 10^8$ s (90\% CL) \cite{re:oscillationsILL}.

Recent proposals leverage new techniques, where the $\bar{n}A$ scattering length, estimated via optical potentials~\cite{re:oscillationsGal}, plays a key role. A new, promising approach foresees the usage of ultracold neutrons (with kinetic energies on the order of 100 neV) trapped in suitable bottles. The experimental sensitivity of such experiments depends on the wall reflectivity for the antineutrons possibly generated following an oscillation in the trap
and on their annihilation rate, which allows their detection. With such an approach an improvement by several orders of magnitude over the current limits of
the $\tau(\bar n n)$ oscillation period could potentially be achieved.

More details on the derivation of the oscillation period with ultra-cold neutrons can be found, for instance, in Ref. \cite{re:fujimodel}. 

The wall reflectivity depends on the potential of the antineutron-nucleus interaction, therefore, as mentioned above, on the antineutron-nucleus scattering length. There are basically two ways to determine such a quantity, both of which invoke hadronic physics:  from a fit of existing X-ray spectroscopy data from antiprotonic atoms, or from the measurement of annihilation cross sections at very low energies, {\textit i.e} in S-wave.  
Data on antiprotonic atoms exist, but unfortunately they do not provide a sufficiently precise estimate at ultra-cold energies. However, very accurate input values are required to properly model the reflection process, in order to properly optimize the bottle materials and shapes.

Measurements of S-wave annihilation and elastic \nbar cross sections are  still missing at present, but could in principle be performed with
backward-produced antineutrons in $\bar p p\rightarrow \bar n n$
CEX reactions, as suggested in  Ref. \cite{re:fujiArxiv, re:proposal} and shortly described in the next Section.

\section{Novel Antineutron beams at the Antiproton Decelerator (AD)}

Producing antineutrons via charge exchange (CEX) to enable the studies outlined above requires slowly extracted antiprotons with momenta $\geq 100$ MeV/$c$. As a figure of merit, 9 MeV/$c$ center-of-mass backward-emitted antineutrons
from CEX from 200 MeV/$c$ antiprotons would be produced with a cross section of $(4.7\pm 1.9)$ $\mu$b/sr \cite{re:fujiArxiv, re:proposal}.

CERN's AD facility currently delivers $4 \times 10^7 \bar{p}$/2 min at 100 MeV/$c$ to the ELENA ring, but still lacks slow extraction. Implementing such a line would demand significant set-up modifications, yet remains, in principle, feasible. If achieved, it could yield spills of $\sim 4 \times 10^4 \bar{p}$, matching LEAR-era intensities.

Past attempts of fast extractions at 300 and 502 MeV/$c$ were successful, and experts deem both slow and high-momentum extraction viable for future AD upgrades, if supported by sufficient community interest.
A key hurdle, however, is the available space: all experimental areas in the AD complex are currently occupied, and new setups would strain the existing infrastructure.

Some ideas for a possible future leverage of the AD facility to complete studies of low-energy interactions at low energy
were described in Ref. \cite{re:proposal}.

\section{Prospects for future Antineutron-Based Experiments}
Some suggestions for simplified set-ups for providing new measurements to clarify  several of the issues discussed above are outlined in the following:

\begin{itemize}
\item Differential elastic and total cross sections measurements:
    \begin{itemize}
    \item \nbar beam: tagged beam with a production (CEX) target and a scatterer;
    \item detector: a compact neutron detector (fiber-based, for instance);
    \item experimental set-up: small equipment, less than 1 m long.
    \end{itemize}

\item Systematic measurements of $\sigma_{ann}(\bar np)$ and $\sigma_{ann}(\bar pp)$ from 700 down to 50 MeV/$c$:

\begin{itemize}
\item non-magnetic set-up;
\item good angular coverage;
\item \nbar beam: produced by CEX, production target needed;
\item reaction target(s): same vessel for LH$_2$ (down to 200 MeV/$c$), gaseous H$_2$ at lower \pbar momenta.
\end{itemize}

\item Meson spectroscopy studies:
\begin{itemize}
\item \nbar beam: intense beam required, slow extraction mandatory;
\item detector: full acceptance magnetic detector with PID capabilities to identify and measure all the emitted particles.
\end{itemize}
\end{itemize}

\section{Conclusions}
Experiments using antineutrons as probes yielded significant results through the 1990s. Yet, limited statistics hampered precision, leaving key questions, and longstanding puzzle unresolved.

Larger datasets, including new samples of $\bar{p}p$ annihilations, are essential for comparisons and resolution of the open issues. No other facility has ever matched, as of today, LEAR's antineutron beam intensity and momentum resolution.

A promising path forward lies in the opportunity of slow extraction of high-energy ($\geq 200$ MeV/$c$) $\bar{p}$ beams from CERN's AD, which could potentially enable a new generation of antineutron experiments revitalizing this research line.

\bibliography{refs.bib}

\end{document}